# Mid-infrared frequency comb spanning an octave based on an Er fiber laser and difference-frequency generation

Fritz Keilmann

LASNIX, Sonnenweg 32, 82152 Berg, Germany

Sergiu Amarie

Max Planck Institute of Quantum Optics, Am Coulombwall 1, 85748 Garching, Germany

Correspondence to: keilmann@lasnix.com

Abstract

We describe a coherent mid-infrared continuum source with 700 $cm^{-1}$ usable bandwidth, readily tuned within 600 - 2500 $cm^{-1}$ (4 - 17 μm) and thus covering much of the infrared "fingerprint" molecular vibration region. It is based on nonlinear frequency conversion in GaSe using a compact commercial 100-fs-pulsed Er fiber laser system providing two amplified near-infrared beams, one of them broadened by a nonlinear optical fiber. The resulting collimated mid-infrared continuum beam of 1 mW quasi-cw power represents a coherent infrared frequency comb with zero carrier-envelope phase, containing about 500,000 modes that are exact multiples of the pulse repetition rate of 40 MHz. The beam's diffraction-limited performance enables long-distance spectroscopic probing as well as maximal focusability for classical and ultraresolving near-field microscopies. Applications are foreseen also in studies of transient chemical phenomena even at ultrafast pump-probe scale, and in high-resolution gas spectroscopy for e.g. breath analysis.

The mid-infrared region spanning the decade of wavelengths from ca. 2.5 to 25 µm covers nearly all fundamental vibration frequencies of molecules and solids. This is why mid-infrared spectroscopy is widely used to analyze and identify chemical compounds. The workhorse instrument for this task is the Fourier-transform infrared (FTIR) spectrometer. It relies on an incoherent thermal light source that covers not only the mid-infrared but also the near-infrared and far-infrared regions.[1] Much stronger brightness over also the full infrared spectrum is available in the incoherent beam lines of many electron synchrotrons.[2] Mid-infrared lasers have been serving important applications in materials processing,[3] sensing,[4] metrology,[5] and communication[6] even though they allow only restricted ranges of step tuning (for example 9-11 µm in the case of the $CO_2$ laser, or 5-7 µm for the CO laser) or of continuous tuning (for example 10 % in the case of quantum cascade lasers). But these lasers cannot support modern extensions of mid-infrared spectroscopy requiring coherent beams with a continuum of frequencies over one or several octaves. Spectroscopic ellipsometry,[7,8] for example, can benefit from coherent beams providing a well-defined, shallow incidence angle with small samples, while spectroscopic near-field microscopy of scattering type (s-SNOM)[9,10] needs diffraction-limited focusability and 10 mW power. Several schemes of attaining mid-infrared coherent continua have been demonstrated, most based on ultrashort pulses from near-infrared lasers feeding nonlinear optical processes of frequency down-conversion.[11-16]

Our system (Fig. 1) is based on difference-frequency generation (DFG) using 100 fs Er fiber laser pulse trains at 1.55 µm wavelength.[17-19] Mixing occurs between 1.55 µm pulses and supercontinuum pulses in the 1.7 - 2.3 µm region which are generated from 1.55 µm pulses by a nonlinear optical fiber built in the laser system. These near-infrared supercontinuum pulses can be spectrally tuned by changing the chirp of the driving 1.55 µm pulses, and this effect tunes the spectral position of the mid-infrared continuum. Its spectrum and power are optimized, at each setting, by small adjustments of the relative pulse delay and of the orientation of the nonlinear crystal.

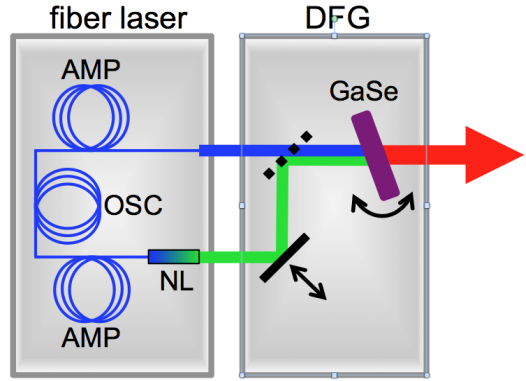

Fig. 1. Schematics of mid-infrared continuum source. The fiber laser system (from toptica.com) contains an oscillator (OSC), two amplifiers (AMP), and a nonlinear optical fiber (NL) producing a red-shifted near-infrared continuum beam (green). The DFG unit (from lasnix.com) combines both free-space output beams at zero pulse delay set by a movable mirror (black arrow) and focuses them on a GaSe crystal which generates the mid-infrared continuum beam (red).

Compared to an earlier setup[18] we achieve about 2x wider usable bandwidth of 700 cm$^{-1}$, and about 40x higher power up to 1 mW. The prime reason for this improvement is the availability of an improved fiber laser system (Toptica mods. FF PRO IR and FF PRO SCIR AMP) which delivers, at 40 MHz repetition rate, 360 mW quasi-cw at 1.55 μm (IR output) and 160 mW quasi-cw in the 1.7 - 2.3 μm region (SCIR output). The optical path length in the DFG unit is held minimal at about 40 cm. A custom polarizing, dichroitic reflector (from ultrafast-innovations.com) allows nearly lossless combining. The second output (not shown in Fig. 1) from the combiner is used to routinely monitor the synchronization of both pulse trains, by two-photon detection. A single lens of 40 mm focal length focuses the combined beams into a 1 mm thick, z-oriented GaSe crystal (from nlo-crystal.tomsk.ru) at 30°–40° off normal depending on the desired wavelength region. The mid-infrared beam is collimated by an Au-coated paraboloidal mirror with 50 mm effective focal length and has a beam diameter of about 10 mm FWHM. Its polarization is essentially linear, with vertical orientation. Its power is measured by a built-in power sensor (Lasnix mod. 511) that has flat response over the 3.7 - 20 μm region and contains a blocking filter for radiation of < 3.5 μm wavelength. Typically the power is between 0.6 and 0.9 mW for appropriate choices of the pump and crystal orientation parameters, independently of where the spectral peak is set in the 800 - 2000 cm$^{-1}$ range (Fig. 2). The spectra were recorded in 2s each by a novel FTIR spectrometer with very compact 15x12 cm$^2$ footprint that can record complete 1 - 36 μm spectra at up to 2 Hz rate (Lasnix mod. L-FTS). Since it allows viewing both the near-infrared and mid-infrared spectra simultaneously, this instrument is highly useful for quick changing (in < 1 min) and optimization of the laser settings, the synchronization, and the crystal orientation.

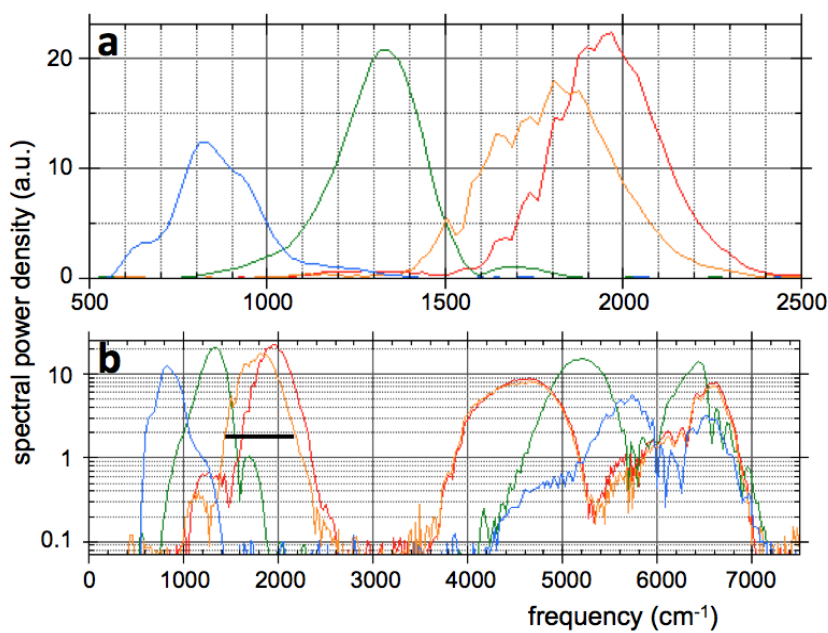

Fig. 2. Power spectra of mid-infrared continuum source at different settings of the fiber laser system (a). The spectra are also shown on an expanded frequency scale for comparing with the generating near-infrared continua (green beam in Fig. 1) which are partly transmitted through the GaSe crystal (b). Both spectra are simultaneously recorded by the L-FTS spectrometer, at about the same signal level by employing a suitable scattering filter.[18] Spectral resolution 25 cm$^{-1}$. The black scale bar represents 700 cm$^{-1}$.

The resulting mid-infrared spectra are smooth and have about 300 cm$^{-1}$ width (FWHM). For any setting, a 700 cm$^{-1}$ wide range (black scale bar) is available for spectroscopy at about a 10% level of the peak spectral power density (ca. 1 μW/cm$^{-1}$ or 1 nW per comb mode). This width amounts to a full octave at the lower end of the tuning range. Note that the corresponding near-infrared continua are somewhat broader. This means that a part of these continua does not contribute to DFG because it is not phase matched. By inspecting the red and orange curves in Fig. 2b we also note that with a given setting of the laser system a substantial shift of the mid-infrared continuum can be obtained by slightly varying the crystal orientation, i.e. the phase matching. In consequence even broader mid-infrared spectra with a usable width up to 1000 cm$^{-1}$ should be obtainable from the same system, by choosing a thinner crystal requiring less stringent phase matching.

Figure 2b demonstrates that the output beam of the DFG source could be used for infrared spectroscopy even over much of the very wide range 500 - 7000 cm$^{-1}$. This may be useful, for example, in near-field microscopy where multispectral contrasts could improve the local chemical recognition. However, if one desires to exploit the frequency-comb property (see below) of the infrared continuum one should remember an important difference: only the mid-infrared parts (600 - 2500 cm$^{-1}$) represent *harmonic* frequency combs where all comb tooth frequencies are exact multiples of the laser's repetition rate, 40 MHz. This is owed to the DFG process which cancels the carrier-envelope offset frequency that might be present in the near-infrared input beams. This offset limits the near-infrared continua (2500 - 7000 cm$^{-1}$), in contrast, to represent frequency combs of a general type. Their use in spectroscopy requires a stabilization of the offset frequency.[20]

In the future, higher power could be obtained trivially once higher-power fiber lasers became available. No restriction would arise from two-photon absorption in GaSe that has been limiting the DFG efficiency in the case of Ti:S laser pumping.[21] Phase matching could be improved by implementing angular dispersion of the near-infrared pulses such that different frequency components propagate in the nonlinear crystal at their individual phase-matching angles (achromatic phase matching).[17,22] But already the present performance enables interesting studies.

One application area of the wide-band infrared coherent continuum source is spectroscopic near-field microscopy even though the attained power is still 10 times short of the ideal value of 10 mW. Convincing near-field spectra were already obtained in exploring polymer fingerprints at 20 nm spatial resolution; for the first time, numerous resonances between 700 and 1800 cm$^{-1}$ could be determined, and this required just two different settings of the coherent continuum source.[23] Another area is spectroscopic ellipsometry which is highly developed in the visible and near infrared where it has become a high-precision analysis method for semiconductor processing; its extension to the mid infrared should now be straightforward and return new insights from specific mid-infrared vibrational and electronic contrasts. A related, unexplored area is complex reflectance spectroscopy in the mid infrared which could now be realized and easily combined with diffraction-limited microscopy; by measuring amplitude and phase spectra this method would determine, for example, complex dielectric functions with much better spatial resolution (and without the need of perfect broadband polarizers) than ellipsometry. Note that precision determination of such spectra is straightforward given the high temporal stability of the 1 mW coherent continuum source, together with the high sensitivity of appropriately broadband HgCdTe detectors of typically < 10 pW noise-equivalent power (for 10 ms integration time). If the beam power of 1 mW were distributed over e.g. 1000 spectral channels the substantial S/N ratio of about 100,000 per channel could be achieved. But really an even much more impressive ratio of S/N = $10^{10}$ applies because in interferometry the detector signal measures the amplitude of the beam in the sample arm while the power in the reference arm can remain at 1 mW. This effect can enable, for example, transmission measurements of highly opaque specimens such as relatively thick, water-containing biological tissues and cells.

The noted sensitivity considerations apply not only to a classical Michelson interferometer but also to a modern dual-comb interferometer.[24,25] This purely time-domain spectrometer requires the superposition of two coherent *harmonic* frequency-comb beams with slightly different repetition frequencies. After its demonstration and first applications[21,26] in the mid-infrared this spectroscopy method has subsequently shown its full potential in the near infrared.[27] Its proven ability of unprecedented accuracy, spectral resolution, and temporal resolution makes its further application in the mid-infrared extremely interesting. Our present frequency-comb source could become instrumental for the analysis of molecular gases as it covers the fundamental vibrations of most molecules. In contrast to common infrared gas analysis which senses mostly only a few vibration-rotational resonances, a broadband assessment of all lines is now possible that enables a complete analysis of a given gas mixture, and this might open new horizons indeed in both environmental probing and in the medical analysis of breath.

A final positive aspect of mid-infrared dual-comb spectroscopy is the ability to record spectra in rapid sequence of e.g. 950 Hz[25] and much higher, thus allowing to follow chemical reactions in real time. Even ultrafast phenomena could be probed with our coherent mid-infrared source. To understand this recall it emits repetitive pulses of about 100 fs duration which can be systematically delayed or advanced in respect to a "pump" pulse which could induce specific excitations of a sample. The pump pulse could be derived from the Er fiber laser and tailored to any wavelength of interest in the visible or near infrared by standard frequency conversion techniques. Successive measurements at varied delay would reveal the complete temporal and spectral evolution of photoinduced reaction dynamics,[28] vibrational dephasing and relaxation, fast conformational changes, and energy or charge transfer events.[29]

In conclusion, we have demonstrated a simple and stable solution for the needs of a number of broadband coherent mid-infrared spectroscopies. Compared with other approaches based on optical parametric oscillators[13] our demonstrated source convinces because of its availability, its large spectral width, its ease of use, its high repetition rate, and also by its small footprint of only 58x44 $cm^2$.

Acknowledgement

We acknowledge helpful dicussions with Marco Marangoni and Albert Schliesser.